\documentclass[aps,prl,twocolumn,superscriptaddress,groupedaddress]{revtex4}  
\usepackage{graphicx}  
\usepackage{dcolumn}   
\usepackage{bm}        
\usepackage{amssymb}   

\begin{document}


\hspace{5.2in} \mbox{FERMILAB-PUB-16-168-E}

\title{Measurement of the forward-backward asymmetries in the production of $\Xi$
and $\Omega$~baryons in $p \bar{p}$ collisions}
\affiliation{LAFEX, Centro Brasileiro de Pesquisas F\'{i}sicas, Rio de Janeiro, RJ 22290, Brazil}
\affiliation{Universidade do Estado do Rio de Janeiro, Rio de Janeiro, RJ 20550, Brazil}
\affiliation{Universidade Federal do ABC, Santo Andr\'e, SP 09210, Brazil}
\affiliation{University of Science and Technology of China, Hefei 230026, People's Republic of China}
\affiliation{Universidad de los Andes, Bogot\'a, 111711, Colombia}
\affiliation{Charles University, Faculty of Mathematics and Physics, Center for Particle Physics, 116 36 Prague 1, Czech Republic}
\affiliation{Czech Technical University in Prague, 116 36 Prague 6, Czech Republic}
\affiliation{Institute of Physics, Academy of Sciences of the Czech Republic, 182 21 Prague, Czech Republic}
\affiliation{Universidad San Francisco de Quito, Quito, Ecuador}
\affiliation{LPC, Universit\'e Blaise Pascal, CNRS/IN2P3, Clermont, F-63178 Aubi\`ere Cedex, France}
\affiliation{LPSC, Universit\'e Joseph Fourier Grenoble 1, CNRS/IN2P3, Institut National Polytechnique de Grenoble, F-38026 Grenoble Cedex, France}
\affiliation{CPPM, Aix-Marseille Universit\'e, CNRS/IN2P3, F-13288 Marseille Cedex 09, France}
\affiliation{LAL, Univ. Paris-Sud, CNRS/IN2P3, Universit\'e Paris-Saclay, F-91898 Orsay Cedex, France}
\affiliation{LPNHE, Universit\'es Paris VI and VII, CNRS/IN2P3, F-75005 Paris, France}
\affiliation{CEA Saclay, Irfu, SPP, F-91191 Gif-Sur-Yvette Cedex, France}
\affiliation{IPHC, Universit\'e de Strasbourg, CNRS/IN2P3, F-67037 Strasbourg, France}
\affiliation{IPNL, Universit\'e Lyon 1, CNRS/IN2P3, F-69622 Villeurbanne Cedex, France and Universit\'e de Lyon, F-69361 Lyon CEDEX 07, France}
\affiliation{III. Physikalisches Institut A, RWTH Aachen University, 52056 Aachen, Germany}
\affiliation{Physikalisches Institut, Universit\"at Freiburg, 79085 Freiburg, Germany}
\affiliation{II. Physikalisches Institut, Georg-August-Universit\"at G\"ottingen, 37073 G\"ottingen, Germany}
\affiliation{Institut f\"ur Physik, Universit\"at Mainz, 55099 Mainz, Germany}
\affiliation{Ludwig-Maximilians-Universit\"at M\"unchen, 80539 M\"unchen, Germany}
\affiliation{Panjab University, Chandigarh 160014, India}
\affiliation{Delhi University, Delhi-110 007, India}
\affiliation{Tata Institute of Fundamental Research, Mumbai-400 005, India}
\affiliation{University College Dublin, Dublin 4, Ireland}
\affiliation{Korea Detector Laboratory, Korea University, Seoul, 02841, Korea}
\affiliation{CINVESTAV, Mexico City 07360, Mexico}
\affiliation{Nikhef, Science Park, 1098 XG Amsterdam, the Netherlands}
\affiliation{Radboud University Nijmegen, 6525 AJ Nijmegen, the Netherlands}
\affiliation{Joint Institute for Nuclear Research, Dubna 141980, Russia}
\affiliation{Institute for Theoretical and Experimental Physics, Moscow 117259, Russia}
\affiliation{Moscow State University, Moscow 119991, Russia}
\affiliation{Institute for High Energy Physics, Protvino, Moscow region 142281, Russia}
\affiliation{Petersburg Nuclear Physics Institute, St. Petersburg 188300, Russia}
\affiliation{Instituci\'{o} Catalana de Recerca i Estudis Avan\c{c}ats (ICREA) and Institut de F\'{i}sica d'Altes Energies (IFAE), 08193 Bellaterra (Barcelona), Spain}
\affiliation{Uppsala University, 751 05 Uppsala, Sweden}
\affiliation{Taras Shevchenko National University of Kyiv, Kiev, 01601, Ukaine}
\affiliation{Lancaster University, Lancaster LA1 4YB, United Kingdom}
\affiliation{Imperial College London, London SW7 2AZ, United Kingdom}
\affiliation{The University of Manchester, Manchester M13 9PL, United Kingdom}
\affiliation{University of Arizona, Tucson, Arizona 85721, USA}
\affiliation{University of California Riverside, Riverside, California 92521, USA}
\affiliation{Florida State University, Tallahassee, Florida 32306, USA}
\affiliation{Fermi National Accelerator Laboratory, Batavia, Illinois 60510, USA}
\affiliation{University of Illinois at Chicago, Chicago, Illinois 60607, USA}
\affiliation{Northern Illinois University, DeKalb, Illinois 60115, USA}
\affiliation{Northwestern University, Evanston, Illinois 60208, USA}
\affiliation{Indiana University, Bloomington, Indiana 47405, USA}
\affiliation{Purdue University Calumet, Hammond, Indiana 46323, USA}
\affiliation{University of Notre Dame, Notre Dame, Indiana 46556, USA}
\affiliation{Iowa State University, Ames, Iowa 50011, USA}
\affiliation{University of Kansas, Lawrence, Kansas 66045, USA}
\affiliation{Louisiana Tech University, Ruston, Louisiana 71272, USA}
\affiliation{Northeastern University, Boston, Massachusetts 02115, USA}
\affiliation{University of Michigan, Ann Arbor, Michigan 48109, USA}
\affiliation{Michigan State University, East Lansing, Michigan 48824, USA}
\affiliation{University of Mississippi, University, Mississippi 38677, USA}
\affiliation{University of Nebraska, Lincoln, Nebraska 68588, USA}
\affiliation{Rutgers University, Piscataway, New Jersey 08855, USA}
\affiliation{Princeton University, Princeton, New Jersey 08544, USA}
\affiliation{State University of New York, Buffalo, New York 14260, USA}
\affiliation{University of Rochester, Rochester, New York 14627, USA}
\affiliation{State University of New York, Stony Brook, New York 11794, USA}
\affiliation{Brookhaven National Laboratory, Upton, New York 11973, USA}
\affiliation{Langston University, Langston, Oklahoma 73050, USA}
\affiliation{University of Oklahoma, Norman, Oklahoma 73019, USA}
\affiliation{Oklahoma State University, Stillwater, Oklahoma 74078, USA}
\affiliation{Oregon State University, Corvallis, Oregon 97331, USA}
\affiliation{Brown University, Providence, Rhode Island 02912, USA}
\affiliation{University of Texas, Arlington, Texas 76019, USA}
\affiliation{Southern Methodist University, Dallas, Texas 75275, USA}
\affiliation{Rice University, Houston, Texas 77005, USA}
\affiliation{University of Virginia, Charlottesville, Virginia 22904, USA}
\affiliation{University of Washington, Seattle, Washington 98195, USA}
\author{V.M.~Abazov} \affiliation{Joint Institute for Nuclear Research, Dubna 141980, Russia}
\author{B.~Abbott} \affiliation{University of Oklahoma, Norman, Oklahoma 73019, USA}
\author{B.S.~Acharya} \affiliation{Tata Institute of Fundamental Research, Mumbai-400 005, India}
\author{M.~Adams} \affiliation{University of Illinois at Chicago, Chicago, Illinois 60607, USA}
\author{T.~Adams} \affiliation{Florida State University, Tallahassee, Florida 32306, USA}
\author{J.P.~Agnew} \affiliation{The University of Manchester, Manchester M13 9PL, United Kingdom}
\author{G.D.~Alexeev} \affiliation{Joint Institute for Nuclear Research, Dubna 141980, Russia}
\author{G.~Alkhazov} \affiliation{Petersburg Nuclear Physics Institute, St. Petersburg 188300, Russia}
\author{A.~Alton$^{a}$} \affiliation{University of Michigan, Ann Arbor, Michigan 48109, USA}
\author{A.~Askew} \affiliation{Florida State University, Tallahassee, Florida 32306, USA}
% ...
%
%
\collaboration{The D0 Collaboration\footnote{with visitors from
$^{a}$Augustana College, Sioux Falls, SD 57197, USA,
$^{b}$The University of Liverpool, Liverpool L69 3BX, UK,
$^{c}$Deutshes Elektronen-Synchrotron (DESY), Notkestrasse 85, Germany,
$^{d}$CONACyT, M-03940 Mexico City, Mexico,
$^{e}$SLAC, Menlo Park, CA 94025, USA,
$^{f}$University College London, London WC1E 6BT, UK,
$^{g}$Centro de Investigacion en Computacion - IPN, CP 07738 Mexico City, Mexico,
$^{h}$Universidade Estadual Paulista, S\~ao Paulo, SP 01140, Brazil,
$^{i}$Karlsruher Institut f\"ur Technologie (KIT) - Steinbuch Centre for Computing (SCC),
D-76128 Karlsruhe, Germany,
$^{j}$Office of Science, U.S. Department of Energy, Washington, D.C. 20585, USA,
$^{k}$American Association for the Advancement of Science, Washington, D.C. 20005, USA,
$^{l}$Kiev Institute for Nuclear Research (KINR), Kyiv 03680, Ukraine,
$^{m}$University of Maryland, College Park, MD 20742, USA,
$^{n}$European Orgnaization for Nuclear Research (CERN), CH-1211 Geneva, Switzerland
and
$^{o}$Purdue University, West Lafayette, IN 47907, USA.
$^{\ddag}$Deceased.
}} \noaffiliation
\vskip 0.25cm
\date{May 11, 2016}

\begin{abstract}
\noindent
We measure the forward-backward asymmetries $A_{\rm FB}$ of charged $\Xi$
and $\Omega$
baryons produced in $p \bar{p}$ collisions
recorded by the
D0 detector at the Fermilab Tevatron collider at 
$\sqrt{s} = 1.96$ TeV as a function of the baryon rapidity $y$.
We find that the asymmetries $A_{\rm FB}$ for charged $\Xi$
and $\Omega$ baryons are consistent with zero within statistical uncertainties.
\end{abstract}

\maketitle

\section{Introduction}
We present a study of the forward-backward asymmetries $A_{\rm FB}$ for the production of charged $\Xi$ and
$\Omega$ baryons 
in $p \bar{p}$ collisions at a center of mass
energy $\sqrt{s} = 1.96$~TeV, recorded by the D0 detector 
at the Fermilab Tevatron collider.

We previously performed a study of $A_{\rm FB}$
for $\Lambda$ and $\bar{\Lambda}$ production~\cite{l}, where $A_{\rm FB}$ is defined as the
relative excess of $\Lambda$ ($\bar{\Lambda}$) baryons produced with
longitudinal momentum $p_z$ in the $p$ ($\bar{p})$ direction.
These results are in agreement
with the observations 
in a wide range of proton collision experiments
that the $\bar{\Lambda}/\Lambda$ production ratio
follows a universal function
of the ``rapidity loss" $y_p - y$ between the
beam proton and the produced $\bar{\Lambda}$ or $\Lambda$ baryon which
does not depend significantly on $\sqrt{s}$
or on the nature of the target $p$, $\bar{p}$, Be, or Pb 
(see Ref. \cite{l} and references therein).
These results support the view that a strange
quark produced directly in the
hard scattering of point-like partons,
or indirectly in the subsequent showering,
can coalesce with a diquark remnant
of the beam particle to produce a $\Lambda$ baryon with a probability
that increases as the rapidity difference between the incoming proton
and outgoing $\Lambda$ baryon decreases.

If this hypothesis is correct, we also expect $A_{\rm FB} > 0$ for
$\Lambda_c$ ($\bar{\Lambda}_c$),
and $\Lambda_b(\bar{\Lambda}_b)$ production
in which a $c$ or $b$ quark can coalesce with a diquark form the proton.
For the $B$ mesons and $\Xi$ and $\Omega$ baryons, we expect
$A_{FB} \approx 0$ since these particles do not share a diquark with the proton.
Previous D0 measurements include $A_{\rm FB}(B^-, B^+)$~\cite{bm} 
and $A_{\rm FB}(\Lambda_b, \bar{\Lambda}_b)$~\cite{lb}.
 
In this article, we present 
measurements of the forward-backward asymmetries
of $\Xi^\mp$ and $\Omega^\mp$ production, where
we use the notation $\Xi^+ \equiv \overline{\Xi^-}$
and $\Omega^+ \equiv \overline{\Omega^-}$. 
The $\Xi^-$ and $\Xi^+$ baryons are defined as ``forward"
if their $p_z$ points in the $p$ or $\bar{p}$ direction, respectively.
The asymmetry $A_{\rm FB}$ is defined as
\begin{eqnarray}
A_{\rm FB} & \equiv & \frac{\sigma_{\rm F}(\Xi^-) - \sigma_{\rm B}(\Xi^-) + \sigma_{\rm F}(\Xi^+) - \sigma_{\rm B}(\Xi^+)}
{\sigma_{\rm F}(\Xi^-) + \sigma_{\rm B}(\Xi^-) + \sigma_{\rm F}(\Xi^+) + \sigma_{\rm B}(\Xi^+)},
\label{lin0}
\end{eqnarray}
where $\sigma_{\rm F}$ and $\sigma_{\rm B}$  are the forward and backward cross sections of $\Xi^-$
or $\Xi^+$ production, and similarly for $\Omega^\mp$ baryons.
The measurements include $\Xi^\mp$ and $\Omega^\mp$ baryons
that are either directly produced or decay products of
heavier hadrons. The measurement strategy for the asymmetry $A_{\rm FB}$ of $\Xi^\mp$
and $\Omega^\mp$ baryons presented here 
closely follows the analysis method used to determine $A_{\rm FB}$ for $\Lambda$ and $\bar{\Lambda}$ baryons
in Ref.~\cite{l}.

\section{Detector and data}

The D0 detector is described in detail 
in Refs.~\cite{run2det, run2muon, layer0, layer0_2}.
The collision region is surrounded by a 
central tracking system that
comprises a silicon microstrip vertex detector and a central fiber
tracker, both located within a $1.9$~T superconducting solenoidal
magnet~\cite{run2det}, surrounded successively by the liquid-argon/uranium
calorimeters, a layer of the muon system~\cite{run2muon}, comprising
drift chambers and scintillation trigger counters, the 
$1.8$~T magnetized iron toroids, and two  
additional muon detector layers after the toroids. 

\begin{figure}[htbp]
\begin{center}
\scalebox{0.38}
{\includegraphics{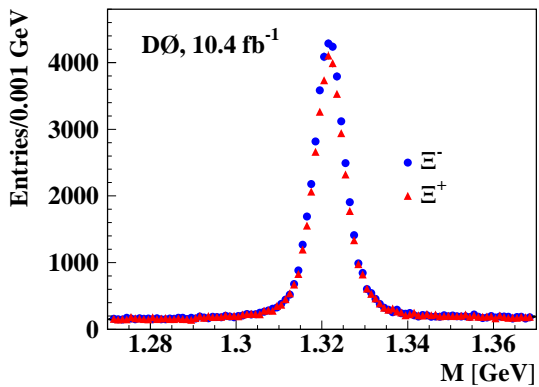}}
\caption{
Invariant mass distributions of 
reconstructed $\Xi^- \rightarrow \Lambda \pi^-$
(circles) and $\Xi^+ \rightarrow \bar{\Lambda} \pi^+$
(triangles) for 
$p \bar{p} \rightarrow \mu \Xi^\mp X$ data.
}
\label{fit_0_1}
\end{center}
\end{figure}

\begin{figure}[htbp]
\begin{center}
\scalebox{0.38}
{\includegraphics{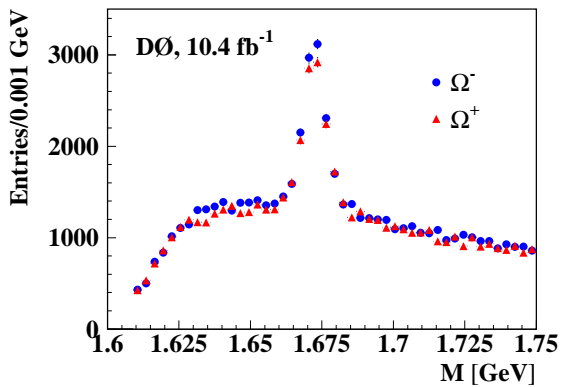}}
\caption{
Invariant mass distributions of reconstructed 
$\Omega^- \rightarrow \Lambda K^-$
(circles) and $\Omega^+ \rightarrow \bar{\Lambda} K^+$
(triangles) for 
$p \bar{p} \rightarrow \mu \Omega^\mp X$ data.
}
\label{m_omega}
\end{center}
\end{figure}

The longitudinal momentum $p_z$ and the rapidity $y \equiv \ln{[(E + p_z)/(E - p_z)]}/2$ 
are both measured with respect to the proton beam direction
in the $p \bar{p}$ center of mass frame where $E$ is the energy of the baryon.
We present results for the full integrated luminosity of $10.4$ fb$^{-1}$, collected from 2002 to 2011,
using two data sets (i) $p \bar{p} \rightarrow \Xi^\mp X$, and
(ii) $p \bar{p} \rightarrow \mu \Xi^\mp X$.
The first data set is unbiased since it is collected 
using a pre-scaled trigger on beam crossing (``zero bias events") or
with a pre-scaled trigger on energy deposited in the forward counters
(``minimum bias events"). 
The second data set is selected with a suite of single muon triggers
which implies that most events contain heavy-quark ($b$ or $c$) decays.
This data set is defined using the same muon triggers and muon selections as 
in Ref.~\cite{dimu_asym3, dimu_asym4}. The muon data provides
a sizable data set that adds additional statistics for the analysis.
For $\Omega$'s there are fewer events, so we only present results for the set
$p \bar{p} \rightarrow \mu \Omega^\mp X$.

We observe $\Xi$ baryons through their decays
$\Xi^-\rightarrow \Lambda \pi^-$ 
and $\Xi^+ \rightarrow \bar{\Lambda} \pi^+$, and
 $\Omega$ baryons through their decays
$\Omega^- \rightarrow \Lambda K^-$ 
and $\Omega^+ \rightarrow \bar{\Lambda} K^+$,
 with $\Lambda \rightarrow p \pi^-$ and $\bar{\Lambda} \rightarrow \bar{p} \pi^+$ in both cases. 
The $\Lambda$ and $\bar{\Lambda}$ candidates are reconstructed
from pairs of oppositely curved tracks with a common vertex ($V^0$). 
Each track is required 
to have a non-zero impact parameter in the transverse plane (IP)
with respect to the $p \bar{p}$ interaction 
vertex with a significance of at least two standard deviations.
The proton (pion) mass is assigned to the daughter track with
larger (smaller) total momentum since the decay
$\Lambda \rightarrow p \pi$ is just above threshold.
The invariant mass of the $(p, \pi^-)$ or $(\bar{p}, \pi^+)$ pair
is required to be in the interval $1.105<M(p\pi)<1.125$~GeV~\cite{l}.
We require $\Lambda$  and $\bar{\Lambda}$ candidates with
$1.5  < p_T < 25$ GeV and pseudorapidity $|\eta| < 2.2$~\footnote{
The pseudorapidity is given by $\eta = -\ln[\tan(\theta/2)]$, 
where $\theta$ is the polar angle with respect to the proton beam direction}, and 
their IP must be non-zero with a significance greater than two standard deviations.

The $\Lambda$ ($\bar{\Lambda}$) candidate is combined with a negatively
(positively) charged-particle track with separation in the transverse
plane from the primary vertex
with significance greater
than three standard deviations and a good vertex with the
$\Lambda$ ($\bar{\Lambda}$) candidate.
This track is assigned the pion mass for $\Xi$'s or the kaon
mass for $\Omega$'s.
The $\Xi^{\mp}$ or $\Omega^{\mp}$ candidates are required to have an IP 
consistent with zero within three standard deviations.
The observed decay lengths in the transverse plane of the 
$\Lambda$ and $\Xi^-$ or $\Omega^-$ (or $\bar{\Lambda}$ and $\Xi^+$ or $\Omega^+$)
are required to be greater than 4 mm. 
The invariant mass for the $\Xi^{\mp}$ candidate is required to be
in the interval $1.2< M(\Lambda\pi)<1.5$~GeV and $1.55<M(\Lambda K)<1.85$~GeV for $\Omega^{\mp}$
candidates. 
The kinematic selections for the $\Xi^{\mp}$ and $\Omega^{\mp}$ candidates are
$p_T > 2.0$~GeV and  $|\eta| < 2.2$.
The pion or kaon track and the two daughter tracks of the $\Lambda$ baryon
are required to be different from any track associated to a muon.
The invariant mass distributions for the
decays $\Xi^- \rightarrow \Lambda \pi^-$ and
$\Xi^+ \rightarrow \bar{\Lambda} \pi^+$
are shown in Fig.~\ref{fit_0_1} and for the
decays $\Omega^- \rightarrow \Lambda K^-$ and
$\Omega^+ \rightarrow \bar{\Lambda} K^+$
in Fig.~\ref{m_omega}.

\section{Raw asymmetries and detector effects}

We obtain the numbers $N_{\rm F}(\Xi^\mp)$ and $N_{\rm B}(\Xi^\mp)$
of reconstructed $\Xi^\mp$ baryons in the forward and backward 
categories in each bin of $|y|$ by counting $\Xi^\mp$
candidates in the signal region, $1.305 < M(\Lambda\pi) < 1.335$~GeV,
and subtracting the counts in two sideband regions, defined by
$1.2775<M(\Lambda\pi)<1.2925$~GeV and $1.3475<M(\Lambda\pi)<1.3625$~GeV.
The signal region for $\Omega^\mp$ candidates is $1.6575<M(\Lambda K)<1.6875$~GeV,
and the sideband regions are $1.630<M(\Lambda K)<1.645$~GeV and $1.700<M(\Lambda K)<1.715$~GeV.

The normalization factor $N$ and the three raw asymmetries $A'_{\rm FB}$, $A'_{\rm NS}$, 
and $A'_\Xi$ are defined by
\begin{eqnarray}
N_{\rm F}(\Xi^-) & \equiv & N (1 + A'_{\rm FB}) (1 - A'_{\rm NS}) (1 + A'_\Xi), \nonumber \\
N_{\rm B}(\Xi^-) & \equiv & N (1 - A'_{\rm FB}) (1 + A'_{\rm NS}) (1 + A'_\Xi), \nonumber \\
N_{\rm F}(\Xi^+) & \equiv & N (1 + A'_{\rm FB}) (1 + A'_{\rm NS}) (1 - A'_\Xi), \nonumber \\
N_{\rm B}(\Xi^+) & \equiv & N (1 - A'_{\rm FB}) (1 - A'_{\rm NS}) (1 - A'_\Xi),
\label{def}
\end{eqnarray}
and similarly for $\Omega$.
The raw asymmetries $A'_{\rm FB}$, $A'_{\rm NS}$, and $A'_\Xi$ 
have contributions from 
the physical asymmetries 
$A_{\rm FB}$, $A_{\rm NS}$, and $A_\Xi$, and from detector effects. 
The forward-backward asymmetry $A_{\rm FB}$ measures the relative excess of  
$\Xi^-$ ($\Xi^+$)
baryons with $p_z$ in the $p$ ($\bar{p}$) direction.
The asymmetry $A_{\rm NS}$ 
is given by the relative excess of the sum of $\Xi^-$ and $\Xi^+$ baryons
with $p_z$
in the $\bar{p}$ beam direction (north) with respect to the $p$ beam direction (south).
The asymmetry $A_\Xi$ is the relative excess of negatively charged 
over positively charged baryons.

The initial $p \bar{p}$ state is invariant with respect to
CP conjugation, which changes the sign of $A_{\rm NS}$ and $A_\Xi$,
while $A_{\rm FB}$ remains unchanged. A non-zero value of $A_{\rm NS}$ or $A_\Xi$ 
would indicate CP violation.

\begin{figure}[htbp]
\begin{center}
\scalebox{0.38}
{\includegraphics{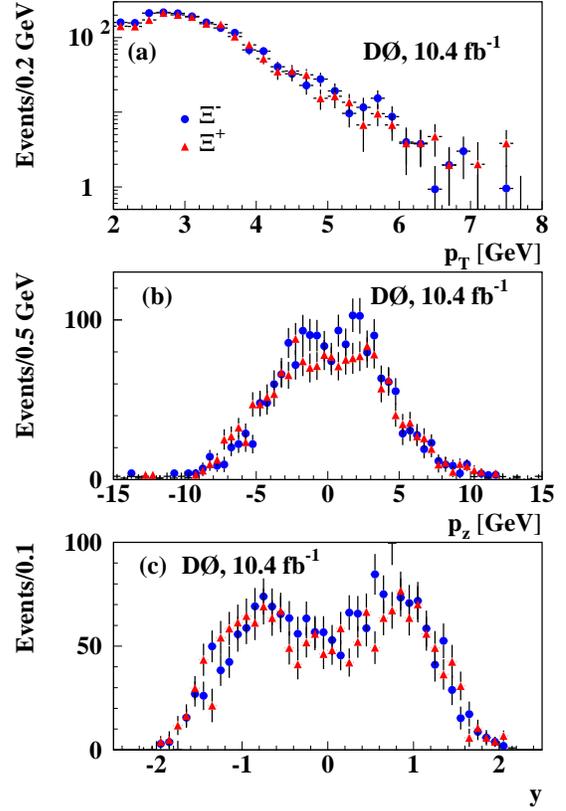}}
\caption{
Distributions of $p_T$, $p_z$, and $y$
of reconstructed $\Xi^-$ (circles) and $\Xi^+$ candidates (triangles)
with $p_T > 2$ GeV, for the minimum bias data sample
$p \bar{p} \rightarrow \Xi^\mp X$.
}
\label{pT_pz_eta_3}
\end{center}
\end{figure}

The asymmetry $A'_{\rm NS}$ is mainly due to differences in the product of the acceptance
and efficiency between the northern hemisphere of the D\O\ detector with respect to
the southern hemisphere. The difference
in reconstruction efficiencies of $\Xi^-$ and $\Xi^+$ baryons caused by the 
different inelastic interaction cross-sections of $p$ and $\bar{p}$
with the detector material creates the additional asymmetry $A'_\Xi$~\cite{l}. 

\begin{figure}[htbp]
\begin{center}
\scalebox{0.35}
{\includegraphics{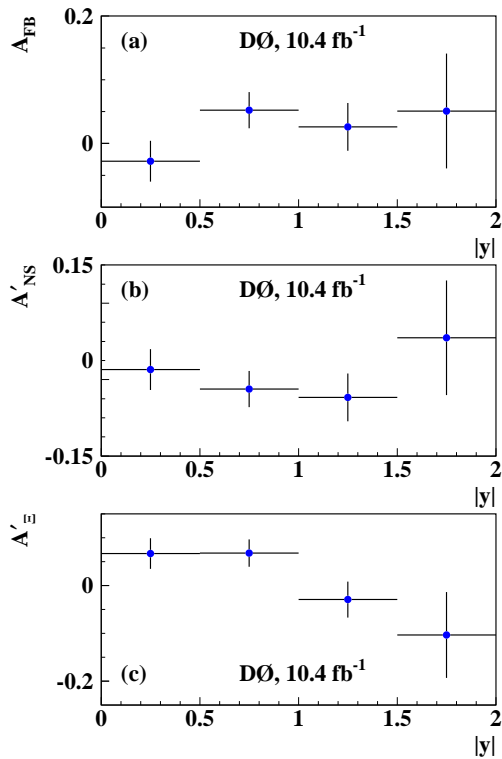}}
\caption{
Asymmetries $A'_{\rm FB} = A_{\rm FB}$,
$A'_{\rm NS}$ and $A'_\Xi$ of reconstructed $\Xi^-$ and $\Xi^+$ candidates
with $p_T > 2$ GeV, as a function of $|y|$,
for the minimum bias data sample
$p \bar{p} \rightarrow \Xi^\mp X$.
The uncertainties are statistical.
}
\label{A_AFB_ANS_3}
\end{center}
\end{figure}

\begin{table*}[htbp]
\caption{\label{results}
Forward-backward asymmetry $A_{\rm FB}$ of
$\Xi^\mp$ baryons with $p_T > 2$ GeV
in minimum bias events,
$p \bar{p} \rightarrow \Xi^\mp X$, and muon events
$p \bar{p} \rightarrow \mu \Xi^\mp X$,
and  $A_{\rm FB}$ of
$\Omega^-$ and $\Omega^+$ baryons with $p_T > 2$ GeV in muon events
$p \bar{p} \rightarrow \mu \Omega^\mp X$.
The first uncertainty is statistical, the second is
systematic due to the detector asymmetry $A'_{\rm NS} A'_\Xi$.
}
\begin{tabular*}{\textwidth}{c @{\extracolsep{\fill}} cccc}
\hline\hline
$|y|$ & $A_{\rm FB} \times 100$ ($\Xi^\mp$, min. bias) & 
$A_{\rm FB} \times 100$ ($\Xi^\mp$, with $\mu$) &
$A_{\rm FB} \times 100$ ($\Omega^\mp$, with $\mu$) \\
\hline
0.0 to 0.5 & $-2.78 \pm 3.20 \pm 0.34$ & $-0.20 \pm  0.72 \pm 0.01$
  & $-3.43 \pm \phantom{1}2.90 \pm 0.13$ \\
0.5 to 1.0 & $\phantom{-}5.23 \pm 2.85 \pm 0.55$  & $-0.13 \pm 0.66 \pm 0.03$
  & $\phantom{-}3.25 \pm \phantom{1}2.78 \pm 0.10$ \\
1.0 to 1.5 & $\phantom{-}2.61\pm 3.75 \pm 0.45 $ & $\phantom{-}1.55\pm0.77 \pm 0.05$
  & $\phantom{-}0.46 \pm \phantom{1}3.52 \pm 0.14$ \\
1.5 to 2.0 & $\phantom{-}5.09 \pm 9.00 \pm 1.64$  & $-1.14\pm 2.05 \pm 0.27$
  & $\phantom{-}5.75 \pm 10.86 \pm 5.70$ \\
\hline\hline
\end{tabular*}
\end{table*}

\begin{figure}[htbp]
\begin{center}
\scalebox{0.38}
{\includegraphics{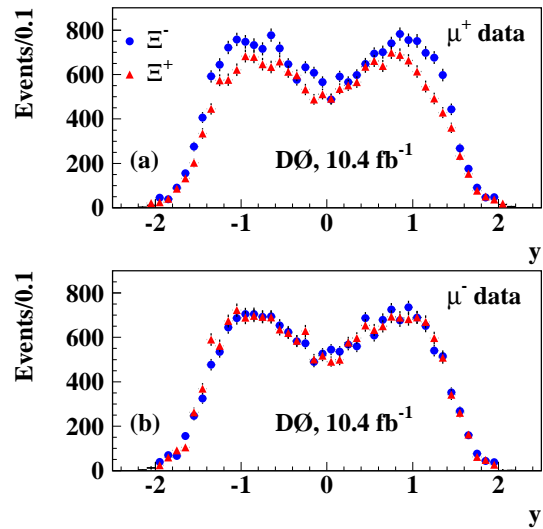}}
\caption{
Distributions of rapidity $y$ for reconstructed $\Xi^-$ (circles) and $\Xi^+$ candidates
(triangles) in events with a (a) positively or (b) negatively charged muon
for $\Xi^{\mp}$ candidates with $p_T > 2$ GeV.
}
\label{y_l_lbar_mup_mum_2_114_4}
\end{center}
\end{figure}

The raw asymmetries 
including terms up to second-order in the asymmetries are given by
\begin{eqnarray}
\lefteqn{A'_{\rm FB}  =  A'_{\rm NS} A'_\Xi } \nonumber \\
& & + \frac{N_{\rm F}(\Xi^-) - N_{\rm B}(\Xi^-) + N_{\rm F}(\Xi^+) - N_{\rm B}(\Xi^+)}
{N_{\rm F}(\Xi^-) + N_{\rm B}(\Xi^-) + N_{\rm F}(\Xi^+) + N_{\rm B}(\Xi^+)},  \label{lin1} \\
\lefteqn{A'_{\rm NS}  =  A'_{\rm FB} A'_\Xi }  \nonumber\\
& & +\frac{-N_{\rm F}(\Xi^-) + N_{\rm B}(\Xi^-) + N_{\rm F}(\Xi^+) - N_{\rm B}(\Xi^+)}
{N_{\rm F}(\Xi^-) + N_{\rm B}(\Xi^-) + N_{\rm F}(\Xi^+) + N_{\rm B}(\Xi^+)},  \label{lin2} \\
\lefteqn{A'_\Xi   = A'_{\rm FB} A'_{\rm NS}  }  \nonumber \\
 & & +\frac{N_{\rm F}(\Xi^-) + N_{\rm B}(\Xi^-) - N_{\rm F}(\Xi^+) - N_{\rm B}(\Xi^+)}
{N_{\rm F}(\Xi^-) + N_{\rm B}(\Xi^-) + N_{\rm F}(\Xi^+) + N_{\rm B}(\Xi^+)}. 
\label{lin3}
\end{eqnarray}

The polarities of the solenoid and toroid magnets
were reversed about once every two weeks during data-taking 
to collect
approximately the same number of events for each of the
four solenoid-toroid polarity combinations. 
We apply weights to equalize the sums of
$\Xi^-$ and $\Xi^+$ candidates reconstructed for each of the
four polarity combinations.
This averaging over magnet polarities
cancels contributions from the detector geometry to $A'_{\rm FB}$ and $A'_\Xi$,
but not to $A'_{\rm NS}$~\cite{l}.

The raw asymmetry $A'_{\rm FB}$
has negligible contributions from detector
effects after averaging over solenoid and toroid magnet polarities.
The raw asymmetries $A'_{\rm NS}$ and $A'_\Xi$ are dominated
by detector effects~\cite{l}. 
The quadratic term $A'_{\rm NS} A'_\Xi$ in Eq. (\ref{lin1})
corrects $A'_{\rm FB}$ for the detector effects $A'_{\rm NS}$ and $A'_\Xi$
on the particle	counts $N_{\rm F}(\Xi^\mp)$ and $N_{\rm B}(\Xi^\mp)$.
We can therefore set $A'_{\rm FB} = A_{\rm FB}$ where 
$A_{\rm FB}$ is defined in Eq. (\ref{lin0}).

\begin{figure}[htbp]
\begin{center}
\scalebox{0.4}
{\includegraphics{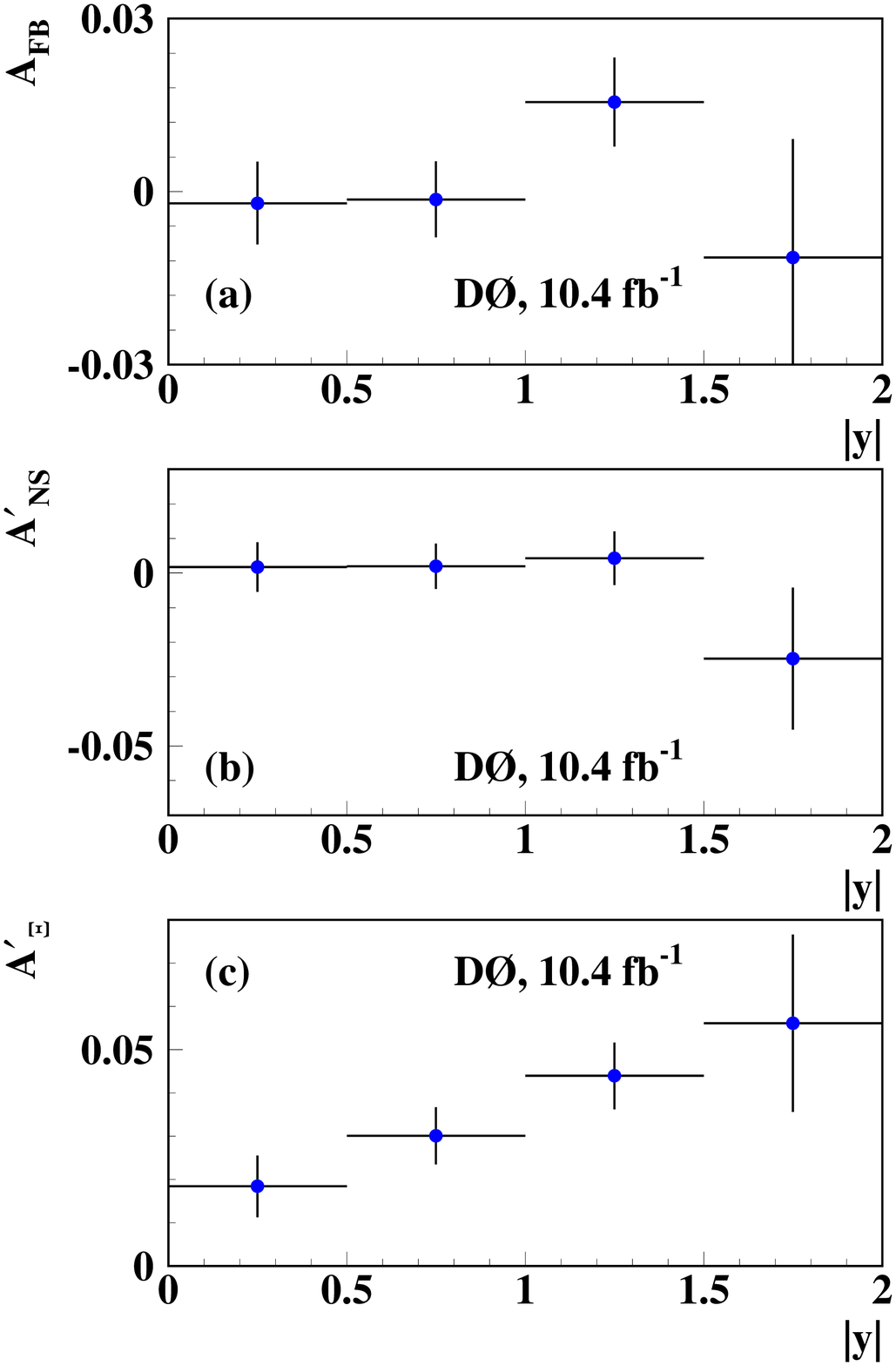}}
\caption{
Asymmetries $A'_{\rm FB} = A_{\rm FB}$,   
$A'_{\rm NS}$ and $A'_\Xi$ of reconstructed $\Xi^-$ and $\Xi^+$ candidates
with $p_T > 2$ GeV, as a function of $|y|$,
for
$p \bar{p} \rightarrow \mu \Xi^\mp X$ events.
The uncertainties are statistical.
}
\label{a_afb_ans_cascade}
\end{center}
\end{figure}

\begin{figure}[htbp]
\begin{center}
\scalebox{0.4}
{\includegraphics{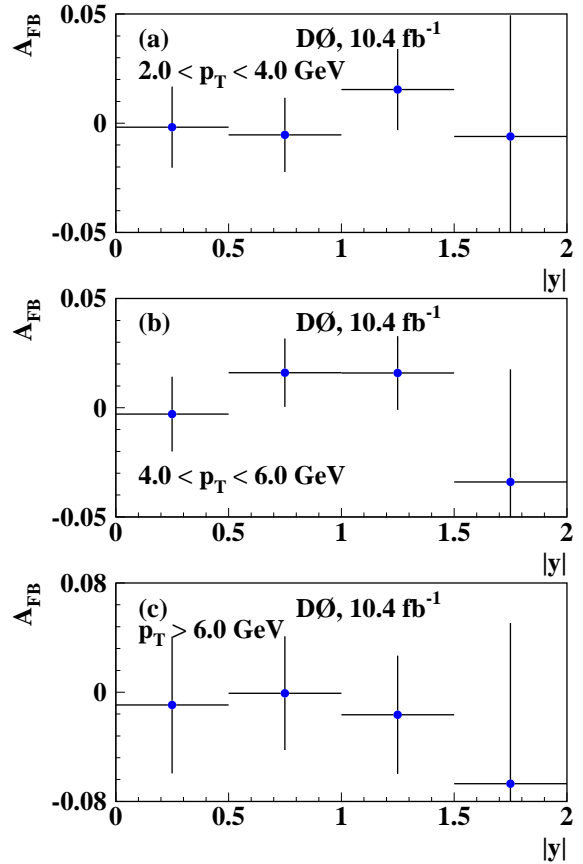}}
\caption{
Asymmetry $A'_{\rm FB} = A_{\rm FB}$ as a function of $|y|$
for $p \bar{p} \rightarrow \mu \Xi^\mp X$ events
with (a) $2.0 < p_T < 4.0$~GeV, (b) $4.0 < p_T < 6.0$~GeV, 
and (c) $p_T > 6.0$ GeV.
The uncertainties are statistical.
}
\label{afb_2_4_6_mup_mum_243}
\end{center}
\end{figure}

\section{Minimum bias sample events $p \bar{p} \rightarrow \Xi^\mp X$}

The minimum bias sample contains $3.7 \times 10^3$ reconstructed
$\Xi^\mp$ candidates with $p_T > 2$ GeV.
Distributions of $p_T$, $p_z$, and $y$ for the  $\Xi^\mp$ candidates are shown in Fig. \ref{pT_pz_eta_3}
and the corresponding raw asymmetries $A'_{\rm FB} = A_{\rm FB}$,
$A'_{\rm NS}$ and $A'_\Xi$ in Fig. \ref{A_AFB_ANS_3}.  
These asymmetries are calculated using Eqs.~\ref{lin1}-\ref{lin3},
neglecting the quadratic terms since they are small compared to the statistical uncertainties.
The correction $A'_{\rm NS} A'_\Xi$ needed to obtain $A'_{\rm FB} = A_{\rm FB}$
 is measured to
be consistent with zero within statistical uncertainties, see Figs.~\ref{A_AFB_ANS_3} (b) and (c).
Thus, we choose not to apply this correction, but rather take
the full measured detector asymmetry $A'_{\rm NS} A'_\Xi$ as the systematic 
uncertainty on the measurement of $A_{\rm FB}$. The
results are summarized in Table \ref{results}.

\section{Muon sample events $p \bar{p} \rightarrow \mu \Xi^\mp X$ 
and $p \bar{p} \rightarrow \mu \Omega^\mp X$}
\label{mu-analysis}

To study the asymmetries using a larger data set, we consider
$p \bar{p} \rightarrow \mu \Xi^\mp X$ and $p \bar{p} \rightarrow \mu \Omega^\mp X$  events
taken from the single muon trigger sample.
Charged particles with transverse momentum in the 
range $1.5 < p_T < 25$ GeV and $|\eta| < 2.2$ are considered as muon candidates.
 Muon candidates are further selected
by matching central tracks with a segment reconstructed
in the muon system and by applying tight quality requirements
aimed at reducing false matching and background
from cosmic rays and beam halo.
To ensure that the muon candidate traverses the detector,
including all three layers of the muon system, we
require either $p_T > 4.2$ GeV or 
$|p_z| > 5.4$ GeV~\cite{dimu_asym3, dimu_asym4}.
The inclusive muon sample contains $2.2 \times 10^9$ reconstructed 
muons and
$7.7 \times 10^4$ reconstructed $\Xi^-$ and $\Xi^+$ candidates with $p_T > 2$ GeV, as well
as  $1.4 \times 10^4$ reconstructed $\Omega^-$ and $\Omega^+$ candidates.

\begin{figure}[htbp]
\begin{center}
\scalebox{0.38}
{\includegraphics{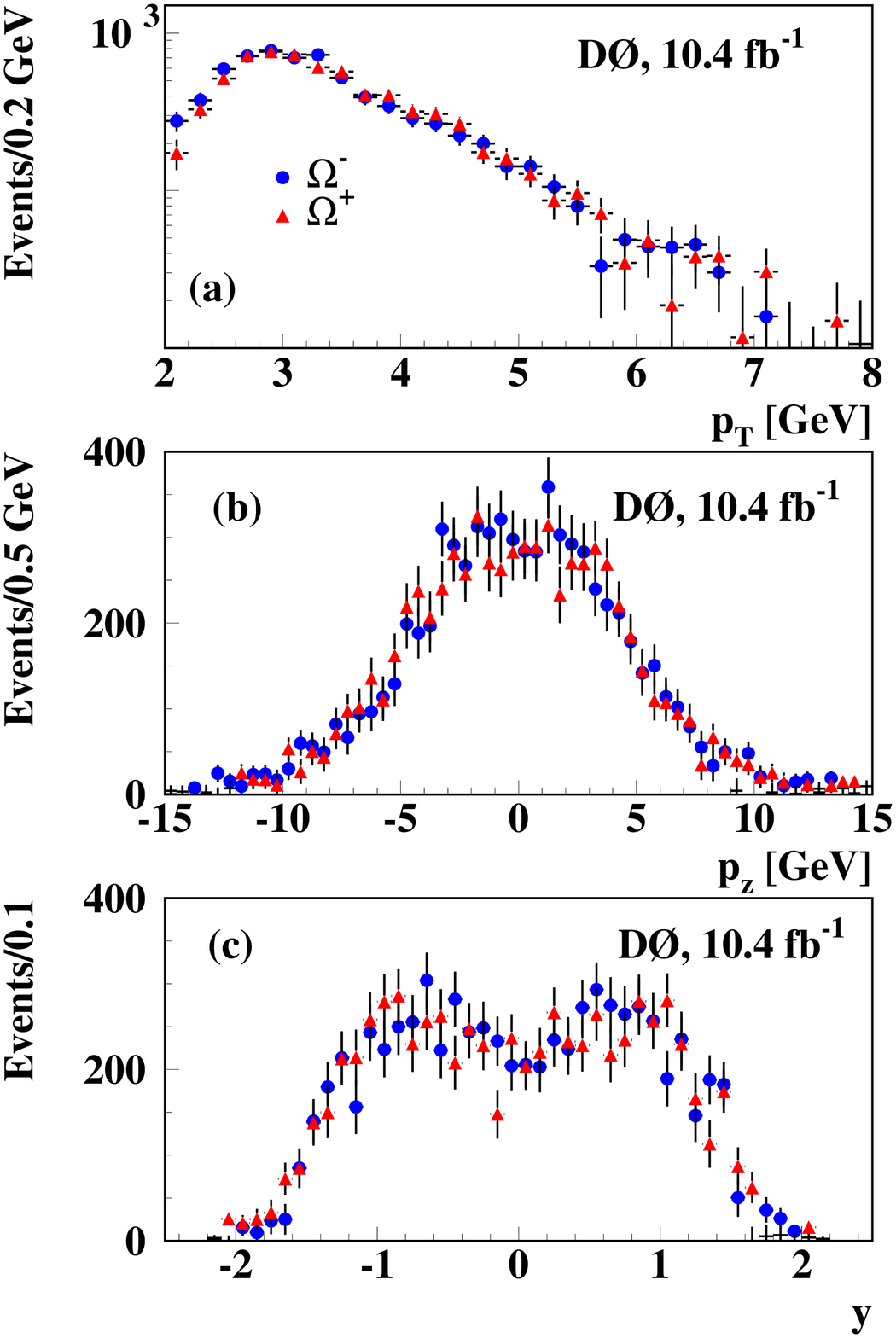}}
\caption{
Distributions of $p_T$, $p_z$, and $y$
of reconstructed $\Omega^-$ (circles) and $\Omega^+$ (triangles)
with $p_T > 2$ GeV, for the data sample
$p \bar{p} \rightarrow \mu \Omega^\mp X$.
}
\label{pt_pz_y_l_lbar_mu_2_152}
\end{center}
\end{figure}

\begin{figure}[htbp]
\begin{center}
\scalebox{0.38}
{\includegraphics{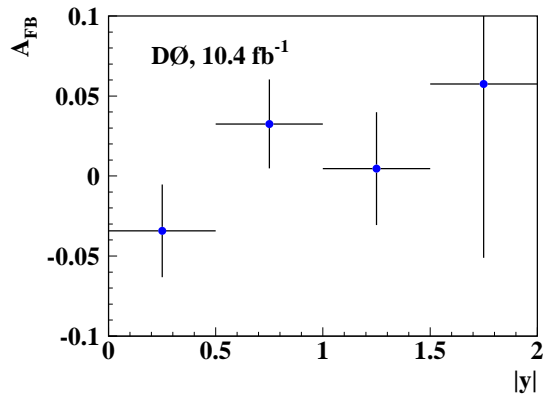}}
\caption{
Asymmetry $A_{\rm FB}$ as a function of $|y|$
for events $p \bar{p} \rightarrow \mu \Omega^\mp X$
for $p_T > 2$ GeV.
The uncertainties are statistical.
}
\label{afb_omega}
\end{center}
\end{figure}

Rapidity distributions for reconstructed $\Xi^-$ and $\Xi^+$ candidates
are shown in Fig.~\ref{y_l_lbar_mup_mum_2_114_4}.
From these distributions 
we observe that (i) the detection efficiency for $\Xi^-$ baryons is larger than for $\Xi^+$
baryons as explained above, and (ii) there are more 
$\Xi^\mp \mu^\pm$ than $\Xi^\mp \mu^\mp$ events.
An example of a process with a correlated $\Xi^- \mu^+$ pair is the decay
$\Xi^0_c \rightarrow \Xi^- \mu^+ X$.
We find that the asymmetry $A'_{\rm FB}$ obtained 
with events containing a $\mu^+$ is consistent with
the corresponding asymmetry with $\mu^-$ within
statistical uncertainties. We therefore
combine the $\mu^+$ and $\mu^-$ samples
to obtain the asymmetries presented in Figs.~\ref{a_afb_ans_cascade} and \ref{afb_2_4_6_mup_mum_243}.

The $p_T$, $p_z$, and $y$ distributions for  $p \bar{p} \rightarrow \mu \Omega^\mp X$ events are shown in
Fig.~\ref{pt_pz_y_l_lbar_mu_2_152}, and the corresponding asymmetry $A_{\rm FB}$ is
presented in Fig.~\ref{afb_omega}. The $\Xi^\mp$ and $\Omega^\mp$ 
asymmetries are summarized in Table \ref{results}.

\section{Conclusions}
We have measured the forward-backward asymmetries $A_{\rm FB}$ in 
$p \bar{p} \rightarrow \Xi^\mp X$,
$p \bar{p} \rightarrow \mu \Xi^\mp X$, and
$p \bar{p} \rightarrow \mu \Omega^\mp X$ events using 
$10.4$~fb$^{-1}$ of integrated luminosity recorded with the D0 detector.
We find that $A_{\rm FB}$ for $\Xi^{\mp}$ and $\Omega^{\mp}$ are 
consistent with zero within uncertainties.

%

We thank the staffs at Fermilab and collaborating institutions,
and acknowledge support from the
Department of Energy and National Science Foundation (United States of America);
Alternative Energies and Atomic Energy Commission and
National Center for Scientific Research/National Institute of Nuclear and Particle Physics  (France);
Ministry of Education and Science of the Russian Federation, 
National Research Center ``Kurchatov Institute" of the Russian Federation, and 
Russian Foundation for Basic Research  (Russia);
National Council for the Development of Science and Technology and
Carlos Chagas Filho Foundation for the Support of Research in the State of Rio de Janeiro (Brazil);
Department of Atomic Energy and Department of Science and Technology (India);
Administrative Department of Science, Technology and Innovation (Colombia);
National Council of Science and Technology (Mexico);
National Research Foundation of Korea (Korea);
Foundation for Fundamental Research on Matter (The Netherlands);
Science and Technology Facilities Council and The Royal Society (United Kingdom);
Ministry of Education, Youth and Sports (Czech Republic);
Bundesministerium f\"{u}r Bildung und Forschung (Federal Ministry of Education and Research) and 
Deutsche Forschungsgemeinschaft (German Research Foundation) (Germany);
Science Foundation Ireland (Ireland);
Swedish Research Council (Sweden);
China Academy of Sciences and National Natural Science Foundation of China (China);
and
Ministry of Education and Science of Ukraine (Ukraine).
%

\end{document}